\documentclass[twocolumn]{pasj01}

\RequirePackage{mathpazo}
\RequirePackage[T1]{fontenc}

\newcounter{author}
\def\authorcount#1#2{\refstepcounter{author}\label{#1}
                     \altaffiltext{\ref{#1}}{#2}}

\Received{$\langle$reception date$\rangle$}
\Accepted{$\langle$acception date$\rangle$}
\Published{$\langle$publication date$\rangle$}

\begin{document}
\SetRunningHead{Y. Tampo et al.}{TCP J21040470+4631129}

\title{First Detection of Two Superoutbursts during Rebrightening Phase of a WZ Sge-type Dwarf Nova: TCP J21040470+4631129}

\author{
        Yusuke~\textsc{Tampo}\altaffilmark{\ref{affil:Kyoto}*},
        Kojiguchi~\textsc{Naoto}\altaffilmark{\ref{affil:Kyoto}},
        Keisuke~\textsc{Isogai}\altaffilmark{\ref{affil:Kyoto}}$^,$\altaffilmark{\ref{affil:KyotoOkayama}},
        Taichi~\textsc{Kato}\altaffilmark{\ref{affil:Kyoto}},
        Mariko~\textsc{Kimura}\altaffilmark{\ref{affil:Kyoto}},
        Yasuyuki~\textsc{Wakamatsu}\altaffilmark{\ref{affil:Kyoto}},
        Daisaku~\textsc{Nogami}\altaffilmark{\ref{affil:Kyoto}},
        Tonny~\textsc{Vanmunster}\altaffilmark{\ref{affil:Van1}}$^,$\altaffilmark{\ref{affil:Van2}},
        Tam\'as~\textsc{Tordai}\altaffilmark{\ref{affil:Polaris}},
        Hidehiko~\textsc{Akazawa}\altaffilmark{\ref{affil:aka}}, 
        Felipe~\textsc{Mugas}\altaffilmark{\ref{affil:IAC_spain}}$^,$\altaffilmark{\ref{affil:ULL}},
        Taku~\textsc{Nishiumi}\altaffilmark{\ref{affil:KSU}}$^,$\altaffilmark{\ref{affil:NAOJ}},
        V\'{\i}ctor J. S.~\textsc{B\'{e}jar}\altaffilmark{\ref{affil:IAC_spain}}$^,$\altaffilmark{\ref{affil:ULL}},
        Kiyoe~\textsc{Kawauchi}\altaffilmark{\ref{affil:EPS_UT}},
        Nicolas~\textsc{Crouzet}\altaffilmark{\ref{affil:ESTEC}},
        Noriharu~\textsc{Watanabe}\altaffilmark{\ref{affil:sokendai}}$^,$\altaffilmark{\ref{affil:NAOJ}}$^,$\altaffilmark{\ref{affil:ABC}},
        N\'{u}ria~\textsc{Casasayas-Barris}\altaffilmark{\ref{affil:IAC_spain}}$^,$\altaffilmark{\ref{affil:ULL}},
        Yuka~\textsc{Terada}\altaffilmark{\ref{affil:UTokyo}},
        Akihiko`\textsc{Fukui}\altaffilmark{\ref{affil:EPS_UT}}$^,$\altaffilmark{\ref{affil:IAC_spain}},
        Norio~\textsc{Narita}\altaffilmark{\ref{affil:ABC}}$^,$\altaffilmark{\ref{affil:IAC_spain}}$^,$\altaffilmark{\ref{affil:NAOJ}}$^,$\altaffilmark{\ref{affil:JST}},
        Enric~\textsc{Palle}\altaffilmark{\ref{affil:IAC_spain}}$^,$\altaffilmark{\ref{affil:ULL}},
        Motohide~\textsc{Tamura}\altaffilmark{\ref{affil:ABC}}$^,$\altaffilmark{\ref{affil:UTokyo}}$^,$\altaffilmark{\ref{affil:NAOJ}},
        Nobuhiko~\textsc{Kusakabe}\altaffilmark{\ref{affil:ABC}}$^,$\altaffilmark{\ref{affil:NAOJ}},
        Roi~\textsc{Alonso}\altaffilmark{\ref{affil:IAC_spain}}$^,$\altaffilmark{\ref{affil:ULL}}, 
        Hiroshi~\textsc{Itoh}\altaffilmark{\ref{affil:Ioh}},
        Kirill~\textsc{Sokolovsky}\altaffilmark{\ref{affil:msu}}$^,$\altaffilmark{\ref{affil:zub2}},
        Brandon~\textsc{McIntyre}\altaffilmark{\ref{affil:msu}},
        Jesse~\textsc{Leahy-McGregor}\altaffilmark{\ref{affil:msu}},
        Stephen~M.~\textsc{Brincat}\altaffilmark{\ref{affil:Brincat}},
        Pavol~A.~\textsc{Dubovsky}\altaffilmark{\ref{affil:Dubovsky}},
        Tom\'a\v{s}~\textsc{Medulka}\altaffilmark{\ref{affil:Dubovsky}},
        Igor~\textsc{Kudzej}\altaffilmark{\ref{affil:Dubovsky}},
        Elena~P.~\textsc{Pavlenko}\altaffilmark{\ref{affil:CrAO}},
        Kirill~A.~\textsc{Antonyuk}\altaffilmark{\ref{affil:CrAO}},
        Nikolaj~V.~\textsc{Pit}\altaffilmark{\ref{affil:CrAO}},
        Oksana~I.~\textsc{Antonyuk}\altaffilmark{\ref{affil:CrAO}},
        Julia~V.~\textsc{Babina}\altaffilmark{\ref{affil:CrAO}},
        Aleksei~V.~\textsc{Baklanov}\altaffilmark{\ref{affil:CrAO}},
        Aleksander~S.~\textsc{Sklyanov}\altaffilmark{\ref{affil:KZN}},
        Alexandra~M.~\textsc{Zubareva} \altaffilmark{\ref{affil:zub}}$^,$\altaffilmark{\ref{affil:zub2}},
        Aleksandr~A.~\textsc{Belinski}\altaffilmark{\ref{affil:zub2}},
        Alexandr~V.~\textsc{Dodin}\altaffilmark{\ref{affil:zub2}},
        Marina~A.~\textsc{Burlak}\altaffilmark{\ref{affil:zub2}},
        Natalia~P.~\textsc{Ikonnikova}\altaffilmark{\ref{affil:zub2}},
        Egor~O.~\textsc{Mishin}\altaffilmark{\ref{affil:zub2}},
        Sergey~A.~\textsc{Potanin}\altaffilmark{\ref{affil:zub2}},
        Ian~\textsc{Miller}\altaffilmark{\ref{affil:Miller}},
        Michael~\textsc{Richmond}\altaffilmark{\ref{affil:rit}},
        Roger~\textsc{D.Pickard}\altaffilmark{\ref{affil:rpc}},
        Charles~\textsc{Galdies}\altaffilmark{\ref{affil:gch}},
        Masanori~\textsc{Mizutani}\altaffilmark{\ref{affil:mzm}},
        Kenneth~\textsc{Menzies}\altaffilmark{\ref{affil:mzk}},
        Geoffrey~\textsc{Stone}\altaffilmark{\ref{affil:sge}} and
        Javier~\textsc{Ruiz}\altaffilmark{\ref{affil:rui1}}$^,$\altaffilmark{\ref{affil:rui2}}$^,$\altaffilmark{\ref{affil:rui3}}
}

\authorcount{affil:Kyoto}{
     Department of Astronomy, Kyoto University, Kyoto 606-8502, Japan}
\email{$^*$tampo@kusastro.kyoto-u.ac.jp}

\authorcount{affil:KyotoOkayama}{
     Okayama Observatory, Kyoto University, 3037-5 Honjo, Kamogatacho,
     Asakuchi, Okayama 719-0232, Japan}

\authorcount{affil:Van1}{
     Center for Backyard Astrophycis Belgium, 
     Walhostraat 1a, B-3401 Landen, Belgium}

\authorcount{affil:Van2}{
    Center for Backyard Astrophycis Extremadura, 
    e-EyE Astronomical Complex, 
    ES-06340 Fregenal de la Sierra, Spain}

\authorcount{affil:Polaris}{
     Polaris Observatory, Hungarian Astronomical Association,
     Laborc utca 2/c, 1037 Budapest, Hungary}

\authorcount{affil:aka}{
    Funao Astronomical Observatory,
    107 Funao, Okayama 710-0261, Japan}
     
\authorcount{affil:IAC_spain}{
    Instituto de Astrof\'isica de Canarias, V\'ia L\'actea s/n,
    E-38205 La Laguna, Tenerife, Spain}
    
\authorcount{affil:ULL}{
    Departamento de Astrof\'isica, Universidad de La Laguna (ULL),
    E-38206 La Laguna, Tenerife, Spain}
    
\authorcount{affil:KSU}{
    Department of Physics, 
    Faculty of Science, Kyoto Sangyo University, 603-8555 Kyoto, Japan}

\authorcount{affil:NAOJ}{
    National Astronomical Observatory of Japan, 
    2-21-1 Osawa, Mitaka, Tokyo 181-8588, Japan}

\authorcount{affil:EPS_UT}{
    Department of Earth and Planetary Science, 
    Graduate School of Science, The University of Tokyo, 7-3-1 Hongo, Bunkyo-ku, Tokyo 113-0033, Japan}

\authorcount{affil:ESTEC}{
    Science Support Office, Directorate of Science, European Space Research and Technology Centre (ESA/ESTEC), Keplerlaan 1, 2201 AZ Noordwijk, The Netherlands}

\authorcount{affil:sokendai}{
    Department of Astronomical Science, The Graduate University for Advanced Studies, SOKENDAI, 2-21-1 Osawa, Mitaka, Tokyo 181-8588, Japan}

\authorcount{affil:ABC}{
    Astrobiology Center, National Institutes of Natural Sciences, 
    2-21-1 Osawa, Mitaka, Tokyo 181-8588, Japan}

\authorcount{affil:UTokyo}{
    Department of Astronomy, Graduate School of Science, 
    The University of Tokyo, 7-3-1 Hongo, Bunkyo-ku, Tokyo 113-0033, Japan}

\authorcount{affil:JST}{
    JST, PRESTO, 2-21-1 Osawa, Mitaka, Tokyo 181-8588, Japan}

\authorcount{affil:Ioh}{
     Variable Star Observers League in Japan (VSOLJ),
     1001-105 Nishiterakata, Hachioji, Tokyo 192-0153, Japan}

\authorcount{affil:msu}{
    Department of Physics and Astronomy,
    Michigan State University, 567 Wilson Rd, East Lansing, MI 48824, USA}

\authorcount{affil:zub2}{
    Sternberg Astronomical Institute, Lomonosov Moscow State University, Universitetsky Ave.,13, Moscow 119992, Russia}

\authorcount{affil:Brincat}{
     Flarestar Observatory, San Gwann SGN 3160, Malta}

\authorcount{affil:Dubovsky}{
     Vihorlat Observatory, Mierova 4, 06601 Humenne, Slovakia}

\authorcount{affil:CrAO}{
     Federal State Budget Scientific Institution ``Crimean Astrophysical
     Observatory of RAS'', Nauchny, 298409, Republic of Crimea}
     
\authorcount{affil:KZN}{     
    FGAOU VO, Kazan (Volga) Federal University, Kazan, Russia}

\authorcount{affil:zub}{
    Institute of Astronomy, Russian Academy of Sciences, Moscow 119017, Russia}

\authorcount{affil:Miller}{
     Furzehill House, Ilston, Swansea, SA2 7LE, UK}
     
\authorcount{affil:rit}{
    Physics Department, Rochester Institute of Technology, Rochester, New York 14623, USA}

\authorcount{affil:rpc}{
    The British Astronomical Association, Variable Star Section (BAA VSS), Burlington House, Piccadilly, London, W1J 0DU, UK}

\authorcount{affil:gch}{
    Institute of Earth Systems, University of Malta, Malta}

\authorcount{affil:mzm}{
    Variable Star Observers League in Japan (VSOLJ),
    Okayama, Japan}
    
\authorcount{affil:mzk}{
    Center for Backyard Astrophysics (Framingham), 318A Potter Road, Framingham, MA01701, USA}
    
\authorcount{affil:sge}{
    Center for Backyard Astrophysics Sierras, 44325 Alder Heights Road, Auberry, CA 93602, USA}

\authorcount{affil:rui1}{
    Observatorio de Cantabria, Ctra. de Rocamundo s/n, Valderredible, 39220, Cantabria, Spain.}
\authorcount{affil:rui2}{
    Instituto de Fisica de Cantabria (CSIC-UC), Avda. Los Castros s/n, 39005, Santander, Spain}
\authorcount{affil:rui3}{
    Agrupacion Astronomica Cantabra, Apartado 573, 39080, Santander, Spain}



\maketitle

\begin{abstract}
We report photometric and spectroscopic observations and analysis of the 2019 superoutburst of TCP J21040470+4631129. 
This object showed a 9-mag superoutburst with early superhumps and ordinary superhumps, which are the features of WZ Sge-type dwarf novae. 
Five rebrightenings were observed after the main superoutburst. 
The spectra during the post-superoutburst stage showed the Balmer, He I and possible sodium doublet features.
The mass ratio is derived as 0.0880(9) from the period of the superhump. 
During the third and fifth rebrightenings, growing superhumps and superoutbursts were observed, which have never been detected during a rebrightening phase among WZ Sge-type dwarf novae with multiple rebrightenings. 
To induce a superoutburst during the brightening phase, the accretion disk was needed to expand beyond the 3:1 resonance radius of the system again after the main superoutburst. 
These peculiar phenomena can be explained by the enhanced viscosity and large radius of the disk suggested by the higher luminosity and the presence of late-stage superhumps during the post-superoutburst stage, plus by more mass supply from the cool mass reservoir and/or from the secondary because of the enhanced mass transfer than those of other WZ Sge-type dwarf novae.

\end{abstract}

\section{Introduction}
\label{sec:1}
Cataclysmic variables (CVs) are the close binary systems made up of a primary white dwarf (WD) and a secondary low-mass star. 
The secondary fills up its Roche lobe, transferring mass into the primary Roche lobe through the inner Lagrangian point $L_1$. 
Dwarf novae (DNe) are a subclass of CVs which possess accretion disks and show recurrent outbursts.
Normal outbursts are typically 2-5-mag brightening for a few days. 
The mechanism of normal outbursts is explained by thermal instability in an accretion disk (\cite{osa74DNmodel}, \cite{mey81DNoutburst}), in that a viscosity jump between neutral and ionized hydrogen triggers a rapid increase of a mass-accretion rate on the WD and the released gravitational energy is observed as an outburst.

SU UMa-type DNe, one subclass of DNe, are characterised by superoutbursts, which are longer and brighter than normal outbursts. 
Superoutbursts come with superhumps, 0.1 - 0.5 mag fluctuations with the period of a few percent longer than the orbital one. 
Superoutbursts are explained as follows; when the disk radius reaches to the 3:1 resonance radius, thermal-tidal instability is triggered and the accretion disk becomes eccentric, which leads to more effective tidal dissipation and brightening \citep{osa89suuma}. 
Superhumps are caused by a precession of the eccentric disk, and the period of superhumps is the synodic period between the precession period and the orbital period of the secondary \citep{whi88tidal, hir90SHexcess}.
\citet{Pdot} proposed that a superoutburst is divided into three stages based on a variation of superhump periods; Stage A has longer superhump periods and growth of superhump amplitudes, Stage B has systematically varying periods and decrease of the amplitudes, and Stage C has shorter periods. 
Also, a term after a superoutburst until reaching its quiescence is refereed as a post-superoutburst stage, and superhumps during this stage are refereed as late-stage superhumps.

WZ Sge-type DNe form a subclass of SU UMa-type DNe, and they show mainly superoutbursts, and seldom normal outbursts due to their low mass-transfer rates (for a review, see \cite{kat15wzsge}). 
The amplitudes of superoutbursts of WZ Sge-type DNe are usually larger than those of SU UMa-type DNe. 
The most outstanding features of WZ Sge-type DNe are early superhumps and rebrightenings. 
Early superhump is a double-wave profile modulation which is observed at an early stage of superoutbursts. 
Since the period of early superhumps is almost same as the orbital one with $\le1\%$ accuracy \citep{ish02wzsgeproc, kat02wzsgeESH}, even if the orbital period were not determined, a WZ Sge-type DN can be studied statistically regarding the periods of early superhumps as the orbital ones \citep{kat15wzsge}. 
When an accretion disk reached the 2:1 resonance radius, tidal instability triggers two-armed pattern in an accretion disk and this pattern is observed as early superhumps \citep{lin79lowqdisk, uem12ESHrecon}. 
Since early superhumps suppress a growth of tidal instability at the 3:1 resonance radius \citep{lub91SHa, osa02wzsgehump, osa03DNoutburst}, early superhumps are observed before Stage A superhumps. 

The second feature of WZ Sge-type DNe is rebrightenings, which are outbursts observed just after a main superoutburst \citep{ric92wzsgedip, osa97egcnc, ham00DNirradiation, kat04egcnc}. 
Mechanism of rebrightenings is yet veiled, though includes some suggestions; mass reservoir model \citep{kat97egcnc, hel01book, osa01egcnc, uem08j1021, iso15ezlyn}, enhanced viscosity model \citep{osa01egcnc, mey15suumareb}, and enhanced mass transfer model \citep{ham00DNirradiation}. 
From the morphology of rebrightening phenomena, based on the repeating times and duration of rebrightenings, WZ Sge-type DNe can be classified into five types; type A : long-duration rebrightening, type B : multiple rebrightenings, type C : single rebrightening, type D : no rebrightening, and type E : double superoutbursts \citep{ima06tss0222, kat15wzsge}. 
This classification may reflect the evolution of binary systems, thus the morphology of rebrightenings could help to understand evolutionary states of DNe \citep{kat15wzsge}. 
The suggested order of evolution is C $\rightarrow$ D $\rightarrow$ A $\rightarrow$ B $\rightarrow$ E.

In this paper, we present observations and analysis of TCP J21040470+4631129 (here after as TCP J2104). 
The superoutburst of TCP J2104 was firstly detected by Hideo Nishimura, Shizuoka-ken, Japan, on three frames using Canon EOS 6D Digital camera + 200-mm f/3.2 lens under the limiting mag = 14.5 on 2019 July 12.490 UT (BJD 2458677.2). 
\footnote{see <http://cbat.eps.harvard.edu/unconf/followups/J21040470+4631129.html> for details.} 
The coordinates of this object are RA: 21:04:04.6784(1) and Dec: +46:31:14.4652(1) (J2000) in the {\it Gaia} Data Release 2 ({\it Gaia} DR2; \cite{GaiaDR2}). 
There is a quiescent counterpart of $G$ = 17.77(5) and $G_{\rm BP}$ $-$ $G_{\rm RP}$ $=$ 0.58(4) at 109.1(1.4)pc, which corresponds to $M_{\rm G} = 12.58(5)$ in absolute magnitude, and the proper motion is 46.7(1) milli-second of arc per year \citep{GaiaDR2, bai18gaia_dist}. 
The orbital period was confirmed as 0.05352(2) d through the spectroscopic observation by \citet{neu19j2104orb}.
TCP J2104 was classified as WZ Sge-type DN since this object showed double-wave early superhumps during the main superoutburst (see Section \ref{sec:3.2}). 
Also, five rebrightenings were observed after the main superoutburst, thus classified as a type B object among WZ Sge-type DNe (see Section \ref{sec:3.4} for details).
Note that ASAS-SN Sky Patrol (ASAS-SN; \cite{ASASSN}) data were contaminated by a nearby star.

Section \ref{sec:2} presents overview of observations of TCP J2104, and Section \ref{sec:3} shows the results of analysis. We discuss the uniqueness and the possible nature of TCP J2104 in Section \ref{sec:4} and give the summary of this paper in Section \ref{sec:5}.

\section{Observations and Analysis}
\label{sec:2}

Time-resolved CCD photometric observations of TCP J2104 were carried out by Variable Star Network (VSNET) collaborations \citep{VSNET}. 
The instruments are summarized in Table E1 \footnote{Table E1 is available only on the online edition as Supporting Information. }, and the logs of photometric observations are listed in Table E2 \footnote{Table E2 is available only on the online edition as Supporting Information. }. 
All the observation epochs are described in Barycentric Julian Date (BJD). 
Note that photometric data include the $g, r, i, z_s, B, V, Rc$ and $Ic$ band filtered and unfiltered data, and the zero-point of the filtered data were adjusted to the data of the $V$ band observations by T. Vanmunster.

The phase dispersion minimization (PDM) method was used for period analysis \citep{PDM}. The 90$\%$ confidence range of $\theta$ statistics by the PDM method was determined following \citet{fer89error, Pdot2}. 
Before period analysis, global trend of the light curve was removed by subtracting a smoothed light curve obtained by locally weighted polynomial regression (LOWESS: \cite{LOWESS}). 
Observed-minus-calculated ($O - C$) diagrams were presented for visualizing period variations, which are sensitive to slight variations of superhump periods.  Note that in this paper, we used 0.0542 d for calculated ($C$).

Also the low resolution spectroscopic data sets were obtained on BJD 2458733.3, 2458746.0, 2458770.0, 2458830.9, 2458831.9, 2458846.0 and 2458851.0 with Kyoto Okayama Optical Low-dispersion Spectrograph with an Integral Field Unit (KOOLS-IFU; \cite{mat19koolsifu}) mounted on the 3.8-m Seimei telescope at Okayama Observatory, Kyoto University 
and on BJD 2458828.2 and 2458832.9 with Transient Double-beam Spectrograph (TDS) mounted on the 2.5-m telescope of Caucasus Mountain Observatory, Sternberg Astronomical Institute, Lomonosov Moscow State University.
The logs of spectroscopic observations are listed in Table E3 \footnote{Table E3 is available only on the online edition as Supporting Information. }.
The wavelength coverage of VPH-Blue of KOOLS-IFU is 4200 - 8000 \AA~and the wavelength resolution is $R = \lambda/\Delta\lambda~\sim$~400 - 600. 
To obtain more detailed data around the H$\alpha$ line, we also performed observations with VPH683. The wavelength coverage of VPH683 of KOOLS-IFU is 5800 - 8000 \AA~and the wavelength resolution is $R \sim$ 2000. 
The wavelength coverages of its blue and red channel dispersers of TDS are 3600 - 5600 \AA~ and 5600 - 7400 \AA~ and the wavelength resolutions are $R \sim$ 1300 and $R \sim$ 2500. 
The data reduction was performed using IRAF in the standard manner (bias subtraction, flat fielding, aperture determination, scattered light subtraction, spectral extraction, wavelength calibration with arc lamps and normalization by the continuum).

\section{Results}
\label{sec:3}

\subsection{overall light curve}
\label{sec:3.1}

\begin{figure*}[tbp]
 \begin{center}
  \includegraphics[width=170mm]{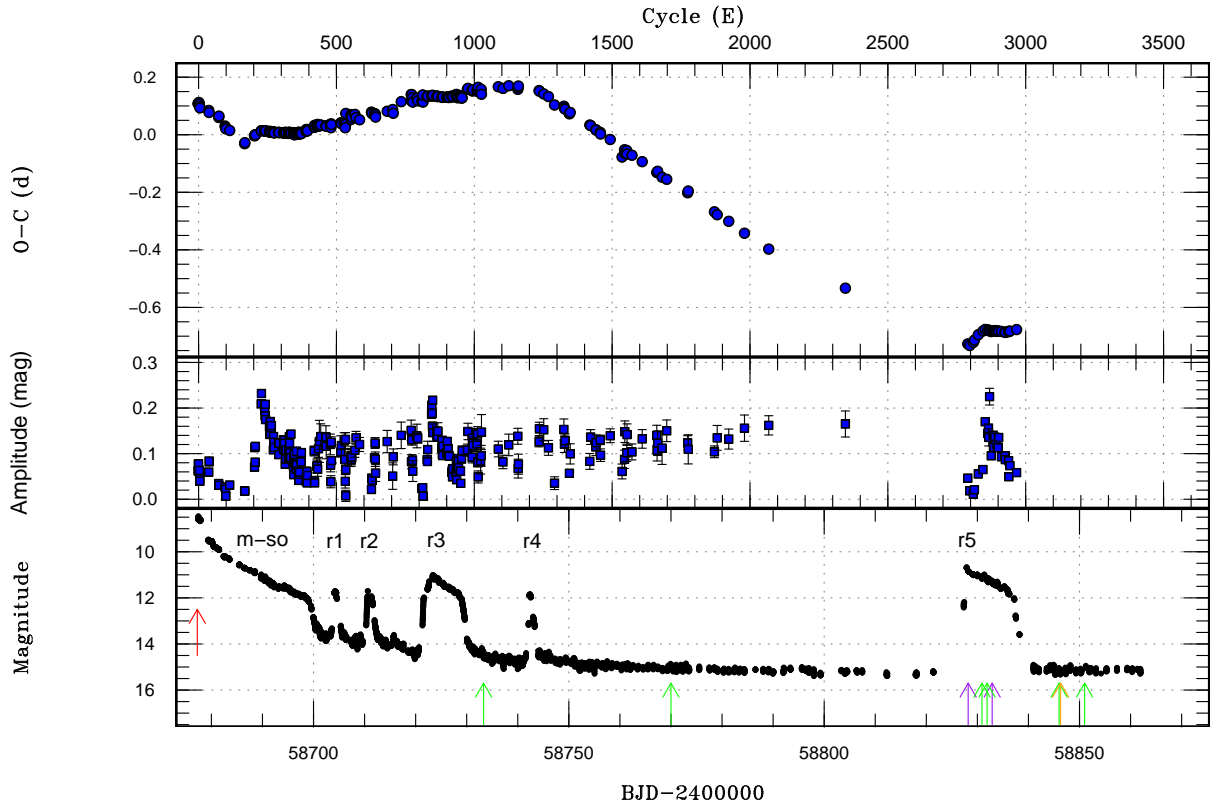}
 \end{center}
 \caption{Top panel : the $O - C$ diagram of TCP J2104. Note that 0.0542 d was used for $C$. Middle panel : the evolution of the superhump amplitudes in the magnitude scale. Bottom panel : the light curve of TCP J2104 during the main superoutburst and rebrightenings. The labels "m-so", "r1", "r2", "r3", "r4" and "r5" mean the main superoutburst and first to fifth rebrightenings. Red arrow presents BJD 2458677.2 when TCP J2104 was firstly detected. Green and purple arrows show the epoch when the spectra were taken (Green : KOOLS-IFU, Purple : TDS). Orange arrow shows the epoch when the photometric data shown in Figure \ref{fig:11} was taken by MuSCAT2 \citep{MuSCAT2}.}
 \label{fig:2}
\end{figure*}

The bottom panel of Figure \ref{fig:2} shows the light curve of TCP J2104 after the detection of the superoutburst \footnote{The over all light curve of pre- and post-detection is presented in Figure E1 on the online edition as Supporting Information.}. The light curve consists of the main superoutburst and five rebrightenings. The main superoutburst lasted for $\sim$24 d from BJD 2458677, reached $\sim$8.5 mag and was $\sim$9 mag brighter at the peak than in quiescence. 
The peak epochs of the first and second rebrightenings were around BJD 2458704.3 and BJD 2458710.0, and their peak magnitudes were $\sim$12 mag. On BJD 2458715, a $>$0.6-mag brightening was observed. 
As AAVSO \footnote{<https://www.aavso.org/>} and ASAS-SN \citep{ASASSN} also detected a brightening on the same epoch, this brightening seems to be real. 
The third rebrightening was detected during BJD 2458721 - 2458279, and reached $\sim$ 11 mag at peak, which was significantly brighter and lasted longer than the first, second and fourth rebrightenings.
The peak epoch of fourth rebrightening was around BJD 2458742.8. After 85 d from the fourth rebrightening, the fifth rebrightening was detected on BJD 2458827 and lasted $\sim$13 d. In this fifth rebrightening, the peak magnitude reached $\sim$11 mag and lasted longer as well.
As of February 2020, its magnitude is $\sim$15.2 and this is still $\sim$3 mag brighter than in quiescence.
 
\citet{}

The $O - C$ diagram of TCP J2104 are presented in the top panel of Figure $\ref{fig:2}$ using a period of 0.0542 d, and a variation of superhump amplitudes in the magnitude scale is shown in the middle panel of Figure $\ref{fig:2}$. Based on the phase changes in the $O - C$ diagram and superhump amplitudes, we regarded BJD 2458677 - 2458686 as the early superhump phase, BJD 2458688 - 2548689 as Stage A, BJD 2458690 - 2458699 as Stage B and later as the post-superoutburst stage, respectively. 
During the third and fifth rebrightening, TCP J2104 showed  the variation of the superhump period and growths of superhump amplitudes. 
Therefore, we determined the superoutburst stages for the third rebrightening: BJD 2458721 - 2458723 as Stage A and BJD 2458723 - 2458729.5 as Stage B, and for the fifth rebrightening as well; BJD 2458827.5 - 2458831 as Stage A and BJD 2458831 - 2458838 as Stage B.

\subsection{early superhumps and ordinary superhumps}
\label{sec:3.2}
\begin{figure}[tbp]
 \begin{center}
    \includegraphics[width=70mm]{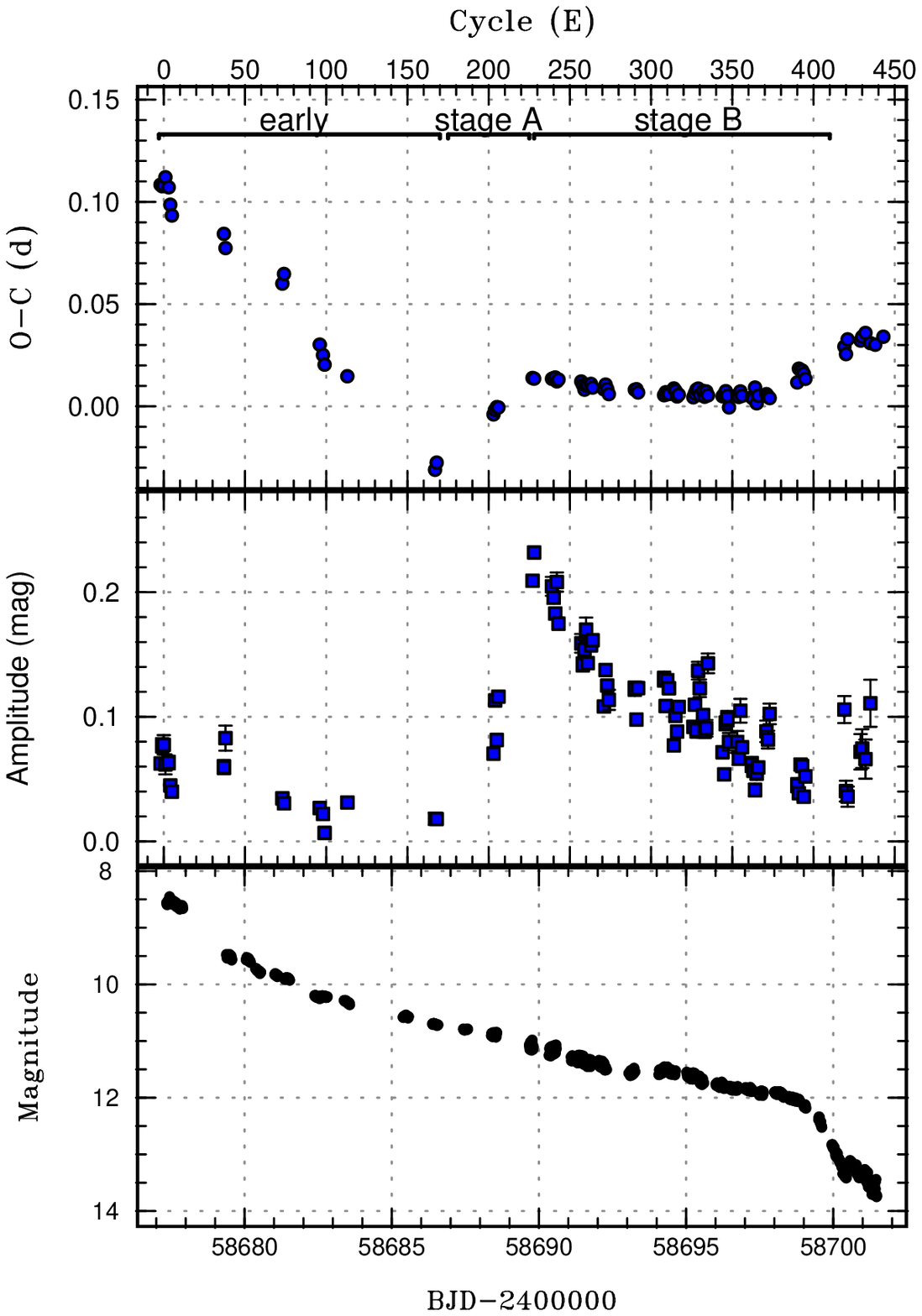}
 \end{center}
 \caption{Top panel : the $O - C$ diagram of TCP J2104 during the main superoutburst. Note that 0.0542 d was used for $C$. Middle panel : the evolution of the superhump amplitudes in the magnitude scale. Bottom panel : the light curve of TCP J2104.}
 \label{fig:14}
\end{figure}

In Figure \ref{fig:14}, the enlarged $O - C$ diagram (upper panel), variation of superhump amplitudes (middle panel) and light curve (lower panel) during the main superoutburst are presented. The evolutions of superhump periods and amplitudes are clearly seen.

\begin{figure}[btp]
 \begin{center}
    \includegraphics[width=80mm]{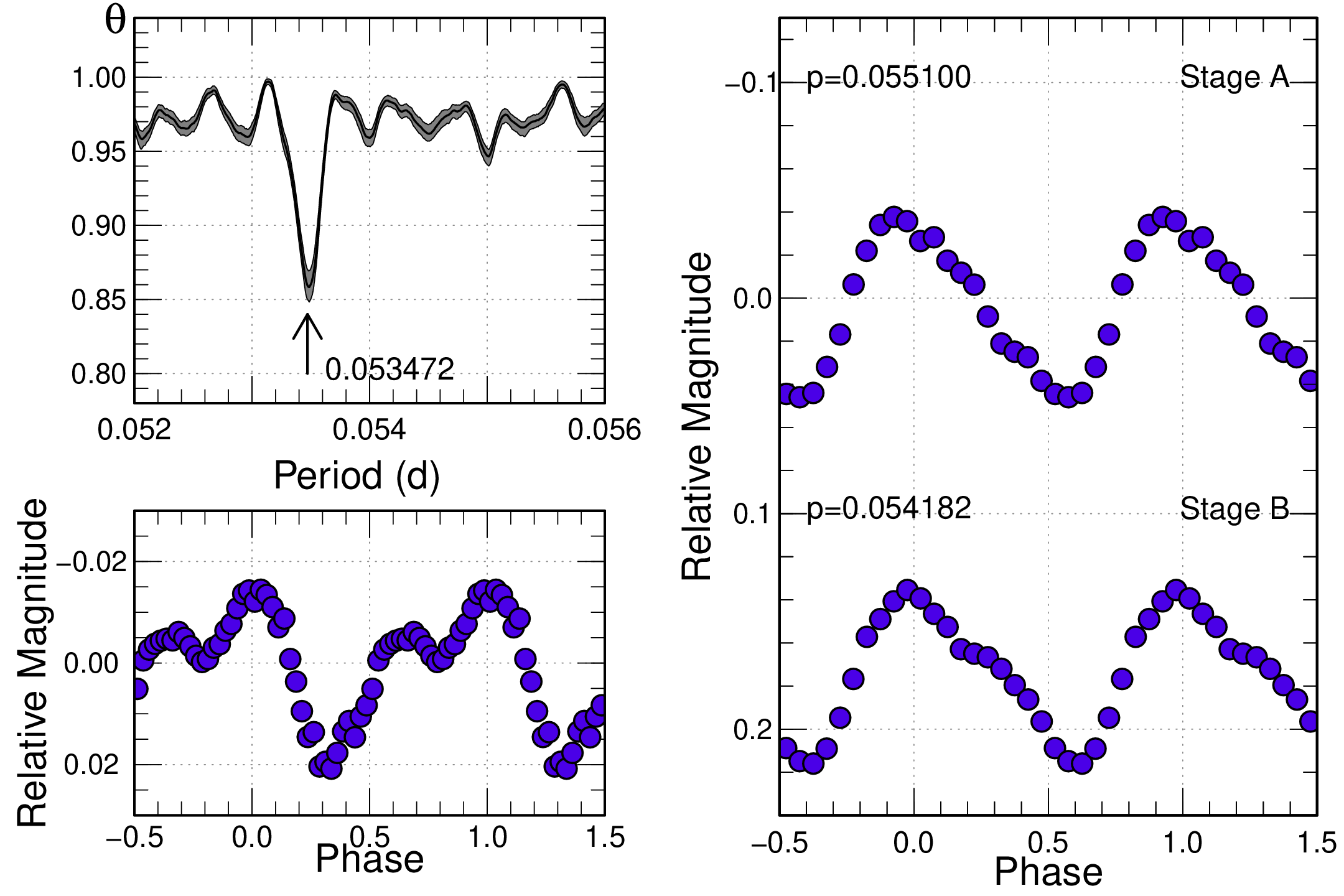}
 \end{center}
 \caption{Upper left panel: $\theta$-diagram of the PDM analysis of early superhumps of TCP J2104 (BJD 2458677 - 2458684). The gray area represents the 90 $\%$ confidence range of $\theta$ statistics by the PDM method \citep{fer89error, Pdot2}. Lower left panel: Phase-averaged profile of early superhumps. Right panel: phase-averaged profiles of the superhumps in Stage A (upper profile) and B (lower profile).}
 \label{fig:3}
\end{figure}

Left panels in Figure \ref{fig:3} show the result of PDM analysis during the early superhump phase (upper panel) and the mean profile of the early superhumps (lower panel). The double-wave variation is clearly seen in the mean profile, and the period of the early superhumps was 0.053472(5) d, which is consistent with the period (0.0535(3) d) reported by \citet{neu19j2104-2}.
This early superhump period is $\sim0.1\%$ shorter than the orbital period reported by \citet{neu19j2104orb} and this slight difference is consistent with other WZ Sge-type DNe \citep{ish02wzsgeproc}. 
Also, Right panel in Figure \ref{fig:3} presents the mean profiles of the superhumps during Stage A (upper profile) and Stage B (lower profile) of the main superoutburst. 
The superhump periods of Stage A and Stage B were 0.0551(1) d and 0.054182(4) d, respectively. 

\subsection{late-stage superhumps and rebrightenings}
\label{sec:3.4}
\begin{figure}[tbp]
 \begin{center}
    \includegraphics[width=60mm]{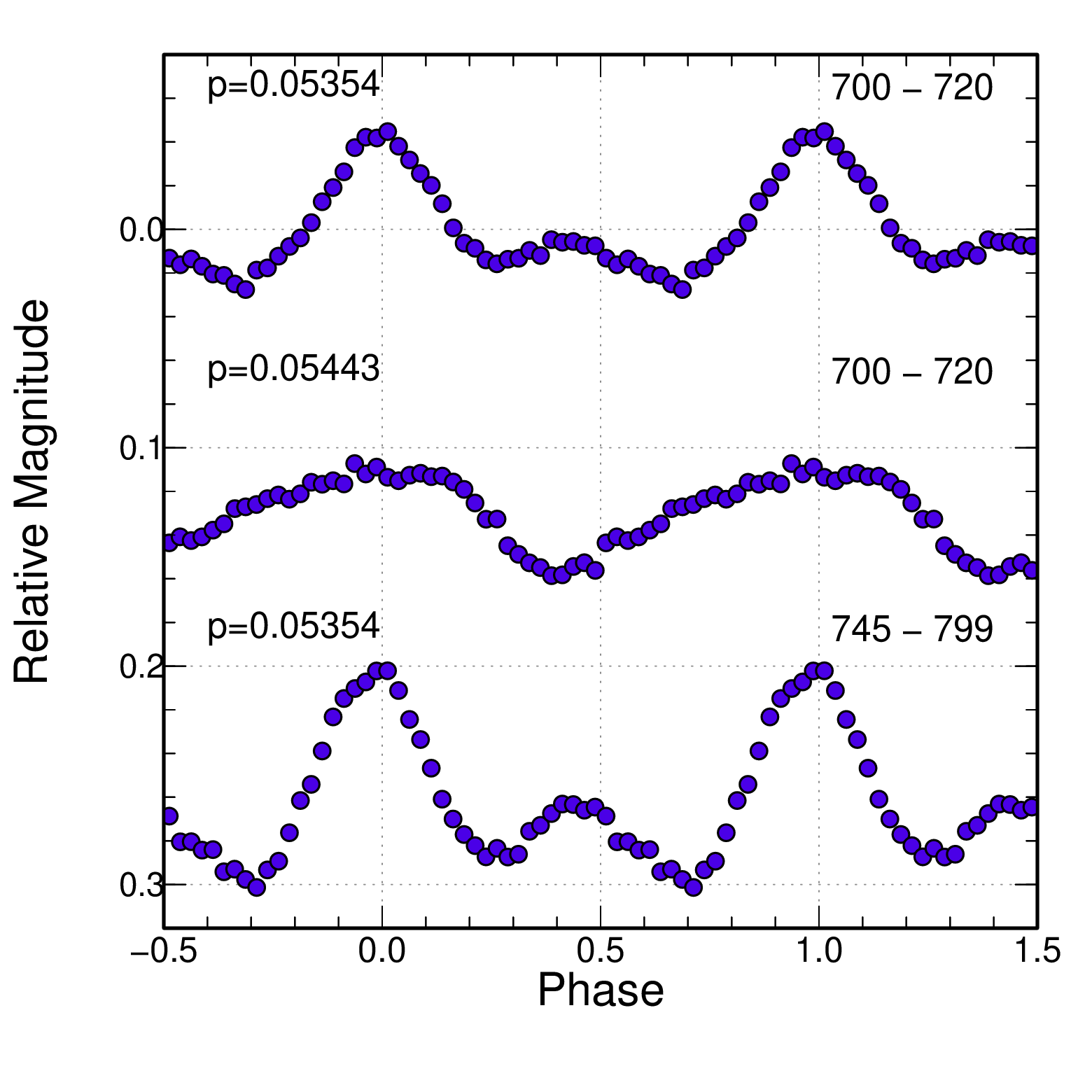}
 \end{center}
 \caption{Phase-averaged profiles of the modulations during BJD 2458700 - 2458720  (upper : 0.053542 d and middle : 0.05443 d) and during BJD 2558745 - 2458799 (lower : 0.05354 d).}
 \label{fig:5}
\end{figure}

PDM analysis of the light curve between the end of the main superoutburst and the start of the third rebrightening suggested two types of variations with periods of 0.053542(6) and 0.05443(1) d, respectively. Figure \ref{fig:5} represent the mean profiles folded with 0.053542 d (upper) and with 0.05443 d (middle).
The former period is close to the orbital period and this can be arose from the emission from a non-axisymmetric disk or a hot spot fixed to the rotational frame of the binary.
Similar double-peak modulations were also detected in ASASSN-14dx \citep{iso19asassn14dx} during the post-superoutburst stage and in quiescence as well. 
The latter period (0.05443 d) seems to be identified as that of the late-stage superhumps since this period is roughly equal to the superhump period during the main superoutburst. 
The presence of the late-stage superhumps implies that the disk still kept the eccentric shape after the main superoutburst \citep{Pdot, kat16rzlmi}. 
Also, the mean profile of the light curve after the fourth rebrightening until the fifth rebrightening is presented in the lower profile of Figure \ref{fig:5}. 
Between the fourth and fifth rebrightening, on the other hand, TCP J2104 only showed variations with the orbital period. 
This indicates that the superhumps ceased after the fourth rebrightening.

\begin{figure*}[tbp]
 \begin{center}
    \includegraphics[width=80mm]{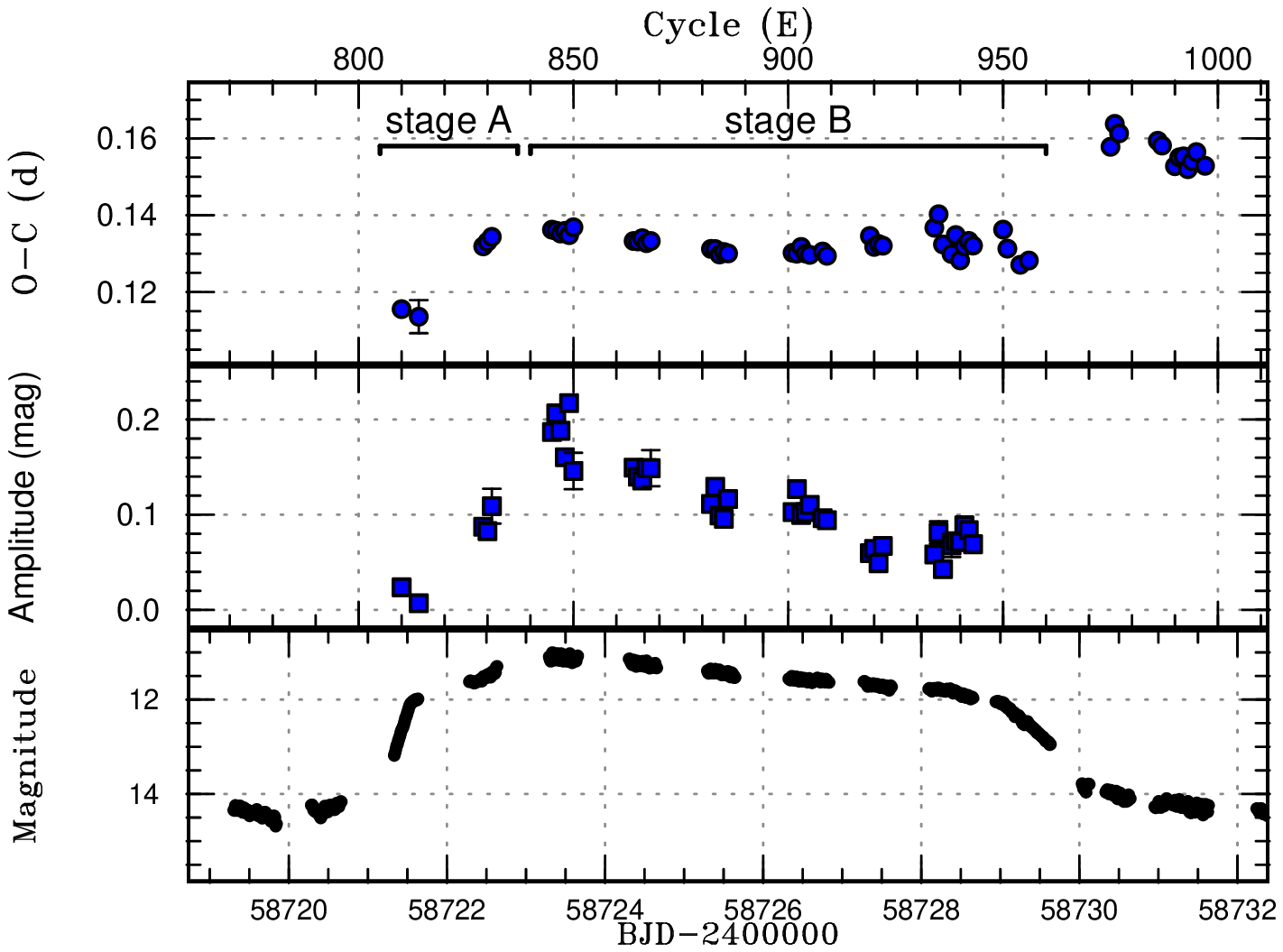}
    \includegraphics[width=80mm]{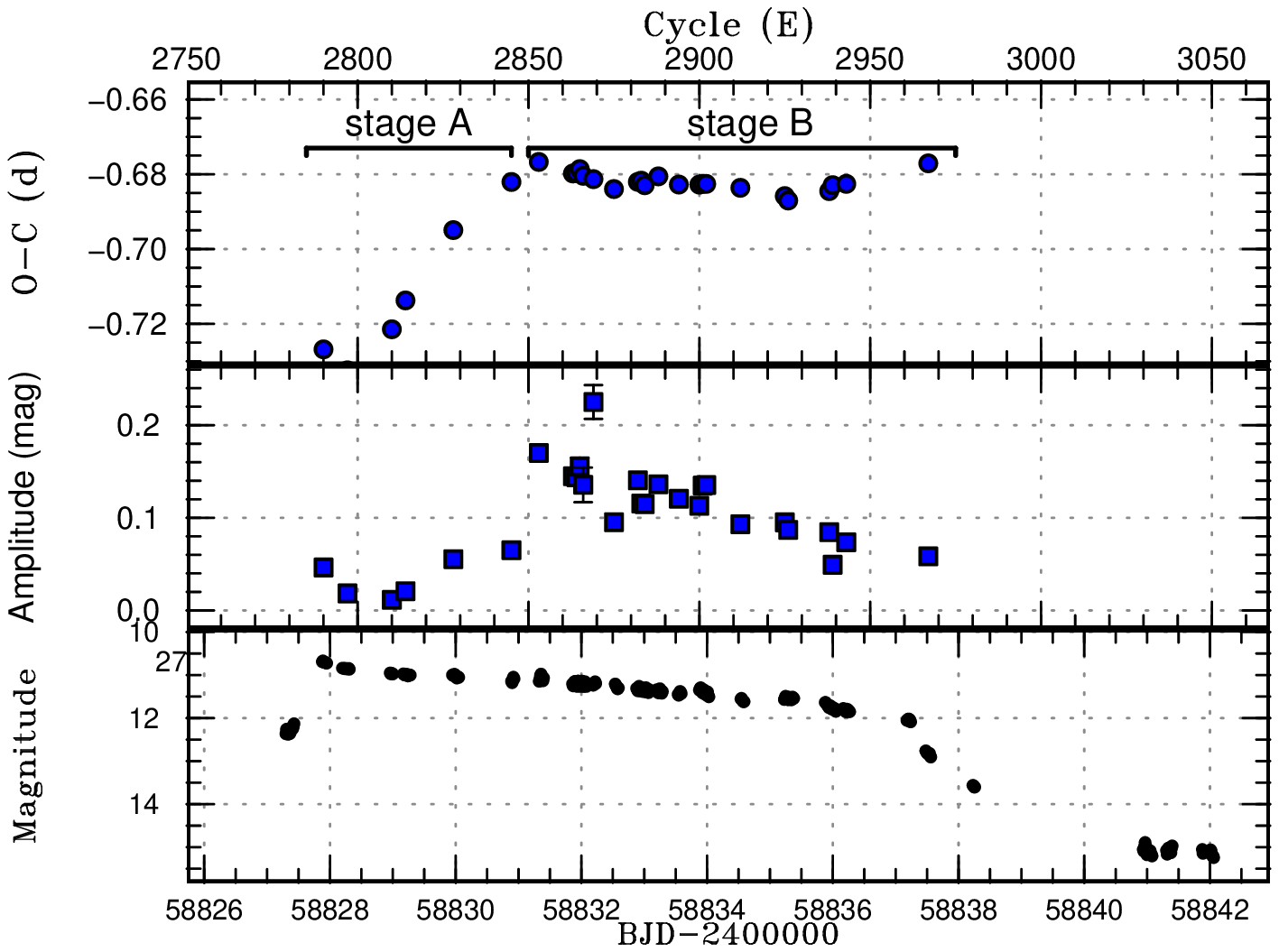}
 \end{center}
 \caption{Left panels are the analysis results of the third rebrightening and right panels are those of the fifth rebrightening. Top panel : the  $O - C$ diagram of TCP J2104 during each rebrightening. Note that 0.0542 d was used for $C$. Middle panel : the evolution of the superhump amplitudes of corresponding phase. Bottom panel : the light curve of TCP J2104 during each rebrightening.}
 \label{fig:7}
\end{figure*}

\begin{figure}[htbp]
 \begin{center}
  \includegraphics[width=60mm]{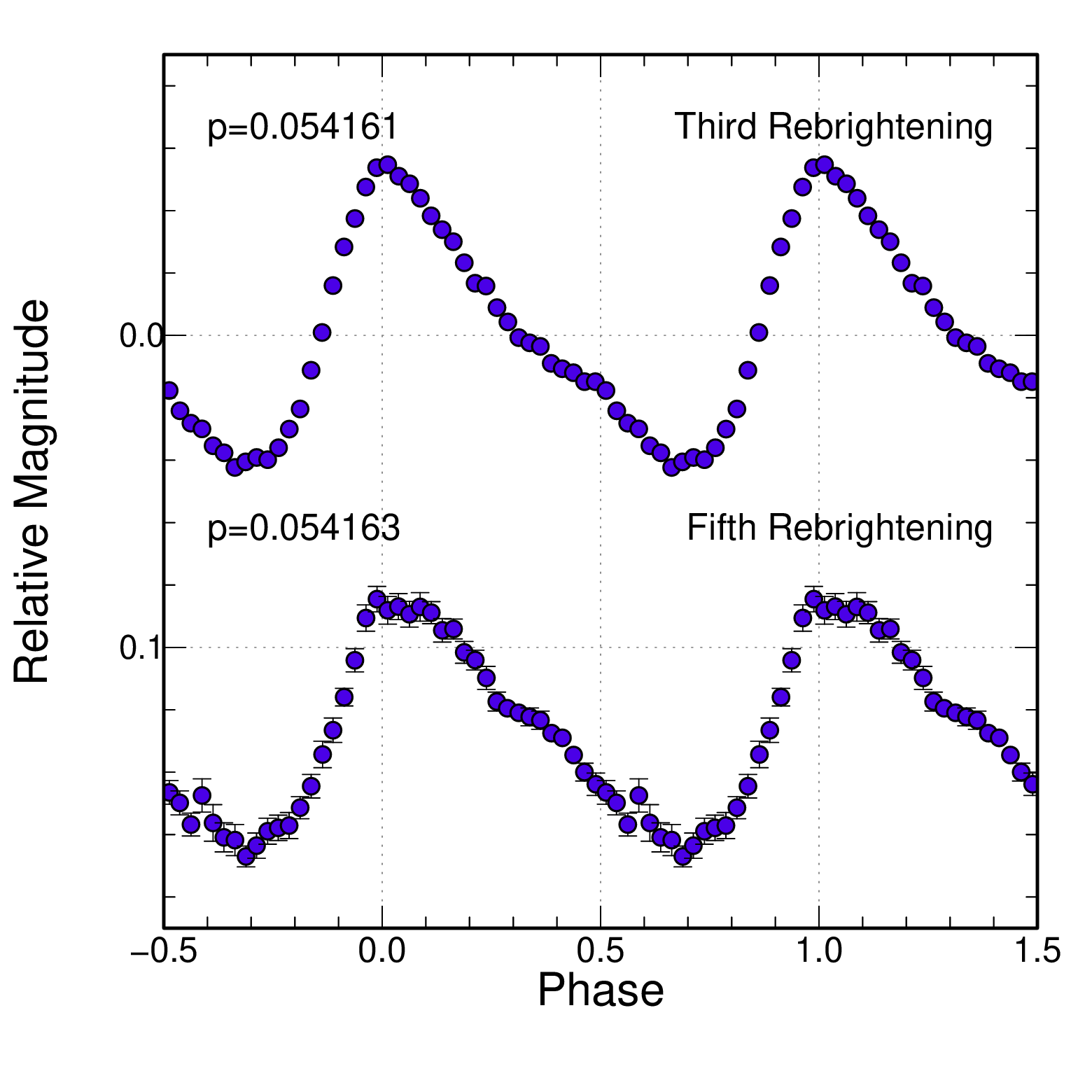}
 \end{center}
 \caption{Phase-averaged profiles during Stage B of the third (upper profile) and fifth (lower profile) rebrightenings.}
 \label{fig:8}
\end{figure}

TCP J2104 showed a total of five rebrightenings after the main superoutburst.
Until this object was observed, detected rebrightenings of all other objects with type B rebrightening (multiple rebrightenings) were only normal outbursts. However in the case of TCP J2104, the period variation and the growth of superhump amplitudes were detected first ever during the rebrightening phase.
The top panels of Figure \ref{fig:7} show the $O - C$ diagrams during the third (left) and fifth (right) rebrightenings, the middle panels do the variations of superhump amplitudes and the bottom panels do the light curves, in which the evolution of superhump periods and amplitudes are clearly seen.
Also Figure \ref{fig:8} presents the mean profiles of superhumps during Stage B of the third (upper profile) and fifth (lower profile) rebrightenings. 
Double-wave modulations were not detected during the third and fifth rebrightenings and the $O - C$ diagrams did not show any possible early superhump stages. 
The peak magnitudes of the third and fifth rebrightenings were 2.5 mag fainter than the main superoutburst (a WZ Sge-type superoutburst) but brighter than the first, second and fourth rebrightenings (normal outbursts). Therefore these features suggest that the third and fifth rebrightenings were SU UMa-type superoutbursts. 
We note that, in the fifth rebrightening the quality of data during Stage A is better than those of the main superoutburst and the third rebrightening, and hence, we used the Stage A superhump period of the fifth rebrightening (0.05530(2) d) for the following discussions.

\subsection{spectroscopic observations}
\label{sec:3.5}
\begin{figure*}[htbp]
 \begin{center}
  \includegraphics[width=170mm]{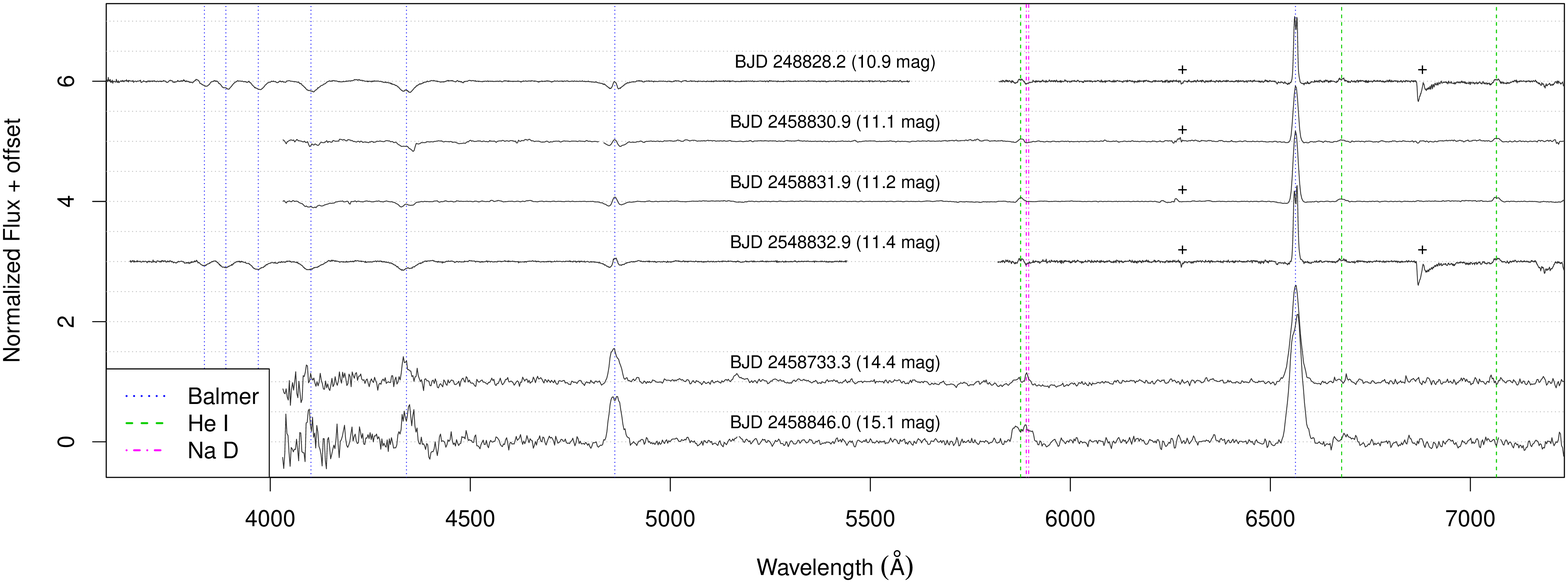}
 \end{center}
 \caption{Low resolution spectra taken on BJD 2458828.2, 2458830.9, 2458831.9, 2458832.9, 2458733.3 and 2458846.0 (upper to lower). Blue lines represents the Balmer features (3835.384, 3889.049, 3970.072, 4101.74, 4340.5, 4861.3 and 6562.8 \AA), green ones do the He I features (5875.65, 6678.1517 and 7065.2 \AA) and pink ones do the Na D features (5889.97 and 5895.94 \AA). Plus marks show the telluric lines (6280 and 6880 \AA).
 All spectra during the fifth rebrightening show H$\alpha$ emission, H$\beta$ emission and absorption lines and He I absorption lines.}
 \label{fig:18}
\end{figure*}

The spectroscopic observations for TCP J2104 were performed during the post-superoutburst stage in Okayama Observatory of Kyoto University and Caucasus Mountain Observatory of Sternberg Astronomical Institute, Lomonosov Moscow State University. 
The six normalized spectra are shown in Figure \ref{fig:18}. 
The upper four spectra were taken during the fifth rebrightening, and lower two spectra were taken before the fourth rebrightening and after the fifth rebrightening. 
All spectra shows the double-peaked Balmer and He I features. 
The spectra during the fifth rebrightening show H$\alpha$ and H$\beta$ absorption lines with a central emission core and also He I emission lines.
This broad absorption line plus superposed narrow emission line at the line center are suggested to be emitted from an optically thick accretion disk and cool gas at the outer disk \citep{cla84rxandktper}. 
The FWHMs of the H$\alpha$ emissions \footnote{Spectra around H$\alpha$ feature are presented in Figure E2 on the online edition as Supporting Information.} are much narrower during the fifth rebrightening ($\sim$500 km/s) than the other epochs ($>$1,000 km/s), and this is essentially consistent with the result reported by \citet{19neuatelj2104}.
Note that the correction of the absorption lines was performed before the estimation of the FWHM of H$\alpha$ emission lines.
In addition, the lower two spectra shows a possible emission line of the sodium doublet (Na D, $\lambda$5889.97/5895.94) with a low S/N, which was also observed during the post-superoutburst stage in GW Lib \citep{vanspa10gwlib} and SSS J122221.7-311525 \citep{neu17j1222}, though this feature is very rare among DNe and could be a noise.

\subsection{light curve before the 2019 superoutburst}
\label{sec:3.6}

{\it Gaia} Survey \citep{gaia} provides the light curve data of TCP J2104 for recent 5 yr \footnote{The light curve from {\it Gaia} survey is presented in Figure E1 on the online edition as Supporting Information.}.
Within $\sim$4.5 yr before the superoutburst in 2019, TCP J2104 got brighter as $\sim$0.5 mag (from 17.92 mag on BJD 2456990 to 17.46 on BJD 2458604, respectively). 
This result is different from the dimming of WZ Sge before the superoutburst \citep{kuu11wzsge}, which was derived from low-quality photometric and visual observations and was less reliable (see \citet{kat15wzsge} for discussion).
This gradual brightening might reflect the accumulation of mass onto the accretion disk, though more samples and, statistical and theoretical analyses are needed.

The other feature of TCP J2104 before the superoutburst in 2019 is the red color according to the Zwicky Transient Facility ($g - r \sim 0.3$; \cite{ztf_cite, ZTF}) \footnote{The light curve from ZTF is presented in Figure E3 on the online edition as Supporting Information.} and {\it Gaia} DR2 (G$_{\rm BP}$ $-$ G$_{\rm RP}$ $=$ 0.58(4); \cite{GaiaDR2}). 
This red color is out of ordinary colors of WZ Sge-type DNe, which usually show blue color spectra dominated by the primary WDs \citep{abr20CVinGaiaHRD}, and actually, G$_{\rm BP}$ $-$ G$_{\rm RP}$ is smaller than 0.4 in most of WZ Sge-type DNe \citep{iso19asassn14dx}.
As TCP J2104 is very close (at 109.2(14) pc; \cite{bai18gaia_dist}) and the dust extinction is almost zero ( E($g-r$) = 0.00(1) at 100 pc for this direction; \cite{gre19dustextinction}), this is the intrinsic color of TCP J2104.
A possible explanation for the red color is the emission from the secondary star. 
If we assumed that only the secondary of TCP J2104 contributes to the $r$ band magnitude of ZTF data \citep{ZTF} in quiescence, the secondary must be $\sim$12.5 mag in absolute magnitude and this might be possible \citep{kni11CVdonor}.

\subsection{short-timescale variations}
\label{sec:3.7}
\begin{figure}[tbp]
 \begin{center}
    \includegraphics[width=70mm]{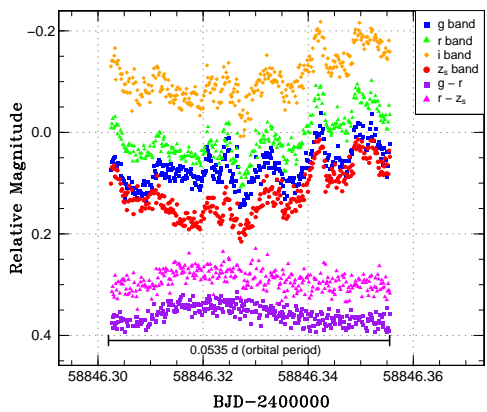}
 \end{center}
 \caption{Highly time-resolved light curves by MuSCAT2 \citep{MuSCAT2} in the $g$ (blue), $r$ (green), $i$ (orange) and $z_s$ (red) band, and $g - r$ (purple) and $r - z_s$ (magenta) on BJD 2458846.3 after the fifth rebrightening. 0.0535 d is the orbital period of TCP J2104.}
 \label{fig:11}
\end{figure}

Highly time-resolved photometric data were taken by MuSCAT (the Multi-color Simultaneous Camera for studying Atmospheres of Transiting planets; \cite{MuSCAT}) mounted on the 1.88-m telescope at Okayama Astrophysical Observatory and MuSCAT2 \citep{MuSCAT2} mounted on the 1.52-m Telescopio Carlos S\'anchez in the Teide Observatory.
Figure \ref{fig:11} shows a light curve of TCP J2104 in the $g, r, i$ and $z_s$ band, and $g - r$ and $r - z_s$ around BJD 2458846.3 after the fifth rebrightening, taken by MuSCAT2. 
Each filtered light curve shows $\sim$0.01 d variations, whose timescales are about one-fifth of orbital or superhump periods.
On the other hand, the $g - r$ and $r - z_s$ colors vary only with orbital or superhump periods.
Similar short-timescale variations were detected in MuSCAT and MuSCAT2 data on the other epochs as well. 
Such kind of variations were observed in ASASSN-18fk \citep{pav19a19fk} and in many systems by \citet{war02DNO} and their following papers. The origin of these kind of modulations are still unclear, and proposed interpretations are a spin period of the white dwarf \citep{pav19a19fk}, flickering due to a variation of mass-transfer rate \citep{sca14flickering} or the travelling waves near the inner edge of the magnetically truncated accretion discs \citep{war02DNO}. 

\section{Discussion}
\label{sec:4}
\subsection{TCP J2104 among WZ Sge-type DNe}
\label{sec:4.1}

\citet{hir90SHexcess} suggested the relation (Equations \ref{eq:1}, \ref{eq:1.1}, \ref{eq:1.2} and \ref{eq:2}) between the mass ratio $q$, orbital period $P_{\rm orb}$ and superhump period $P_{\rm sh}$, using the dimensionless radius $r$ normalized by the binary separation $A$,

\begin{equation}
    \label{eq:1}
    1 - \frac{P_{\rm orb}}{P_{\rm sh}} = Q(q) \times R(r).
\end{equation}    

\noindent
The dependence on $q$ and $r$ are

\begin{equation}
    \label{eq:1.1}
     Q(q) = \frac{1}{2}\frac{q}{\sqrt{1 + q}}
\end{equation}   

\noindent
and 

\begin{equation}
    \label{eq:1.2}
    R(r) = \frac{1}{2} \ \sqrt{r} \ b^{(1)}_{3/2}(r)
\end{equation}   

\noindent
where $\frac{1}{2} b^{(j)}_{s/2}(r)$ is the Laplace coefficient

\begin{equation}
    \label{eq:2}
    \frac{1}{2} b^{(j)}_{s/2}(r) = \frac{1}{2\pi}\int^{2\pi}_0  \frac{\cos (j\phi) d\phi}{(1 + r^2 -2r\cos \phi)^{s/2}}.
\end{equation}

\noindent
During Stage A, superhumps are growing at the 3:1 resonance radius. Thus the mass ratio is obtained by substituting Equation \ref{eq:3} and \ref{eq:3.1} for $r$ and $P_{\rm sh}$ in Equation \ref{eq:1} \citep{kat13qfromstageA},

\begin{equation}
    \label{eq:3}
    1 - \frac{P_{\rm orb}}{P_{\rm Stage~A~sh}} = Q(q) \times R(r_{3:1})
\end{equation}

\begin{equation}
    \label{eq:3.1}
    r_{3:1} = 3^{-2/3} \ (1+q)^{-1/3}.
\end{equation}

\begin{figure}[tbp]
 \begin{center}
  \includegraphics[width=80mm]{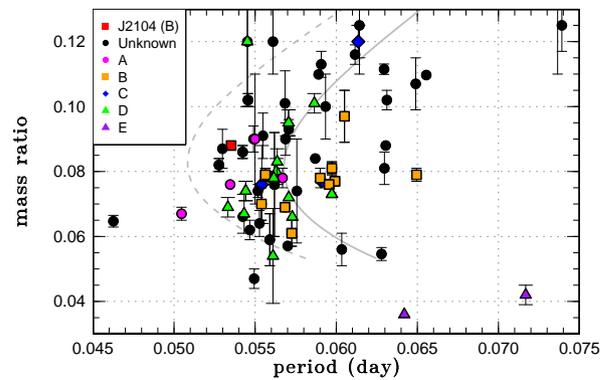}
 \end{center}
 \caption{Relations between orbital periods or early superhump periods $P_{\rm orb}$ and mass ratios $q$ of WZ Sge-type DNe. Red square corresponds to TCP J2104, and other markers represent WZ Sge-type DNe form \citet{Pdot9}. Shapes and colors of markers distinguish the type of rebrightenings; black circles : type unknown, magenta circles : type A (long-duration rebrightening), orange squares : type B (multiple rebrightenings), blue diamonds : type C (single rebrightening), green triangles : type D (no rebrightening), and purple triangles : type E (double superoutbursts). Solid and dashed line represents the theoretical model and best fit model of the evolutionary track of DNe \citep{kni11CVdonor}.}
 \label{fig:12} 
\end{figure}

\noindent
The mass ratio for TCP J2104 was determined as 0.0880(9) using this method.
Figure \ref{fig:12} shows the relations between mass ratios and orbital periods or early superhump periods for TCP J2104 and other WZ Sge-type DNe from \citet{Pdot9}, also showing the theoretical evolution track of DNe assuming mass of a primary WD as 0.75$M_\odot$ \citep{kni11CVdonor}.
TCP J2104 is located around the period minimum, which is consistent with other WZ Sge-type DNe with multiple rebrightenings (type B rebrightening).

Note that, in contrast, \citet{neu19j2104orb} estimated the mass ratio to be $\sim$0.1 using the empirical relation proposed by \citet{pat05SH} and adopting the superhump period as 0.0547(3) d.
Their superhump period might be that during Stage A of the main superoutburst, even though that during Stage B should be adopted as the superhump period for this relation. 
Using our superhump period during Stage B and this relation, the mass ratio of TCP J2104 is estimated to be $\sim$0.062.
Since this empirical relation was obtained from the mass ratio of just 12 DNe which are mostly SU UMa-type and would not be suitable for such a peculiar WZ Sge-type DN, the accurate mass ratio of TCP J2104 seems to be 0.0880(9) estimated by the method proposed by \citet{kat13qfromstageA} which reflects the possible accretion disk physics at the 3:1 resonance radius.

\begin{figure}[tbp]
 \begin{center}
    \includegraphics[width=40mm]{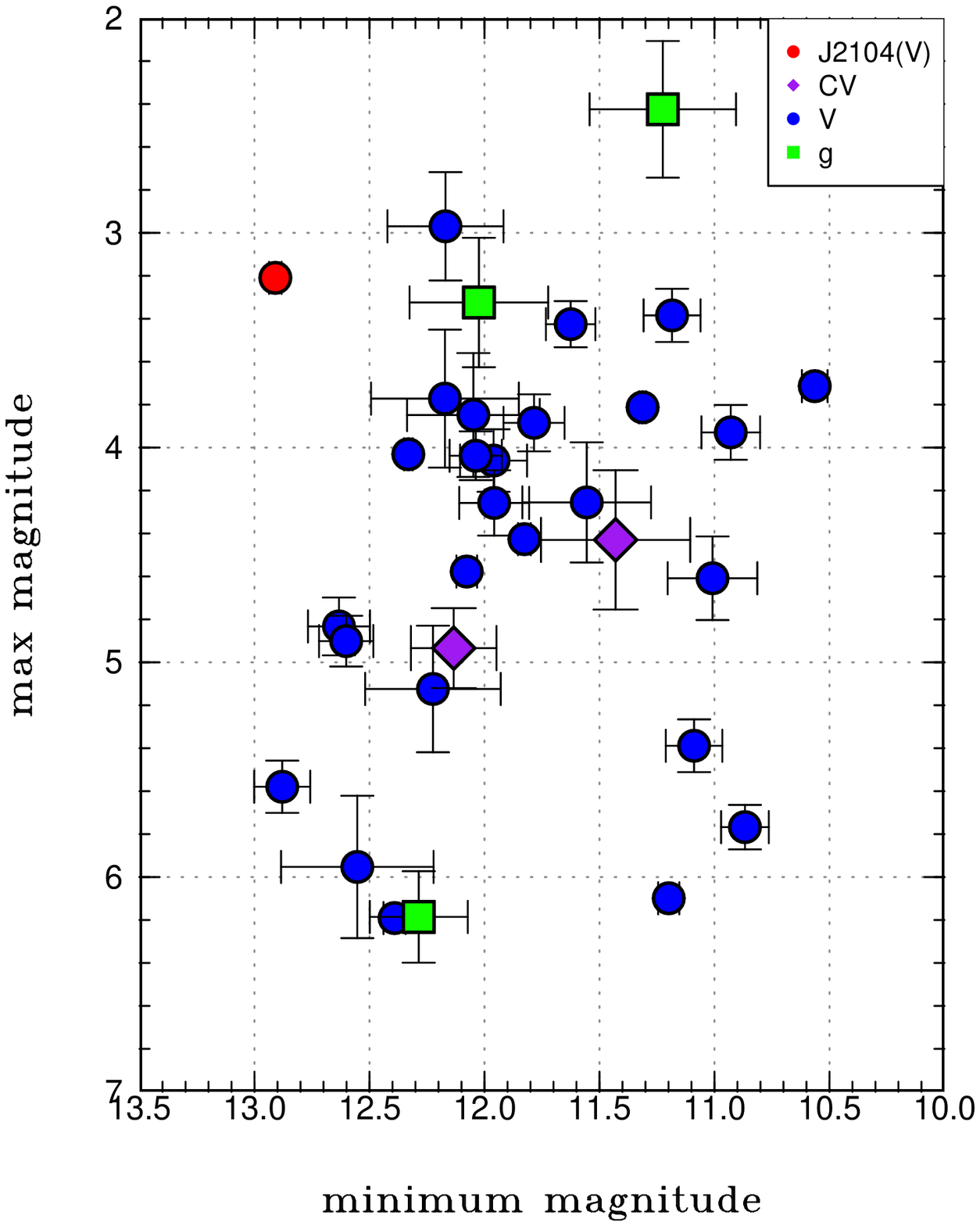}
    \includegraphics[width=40mm]{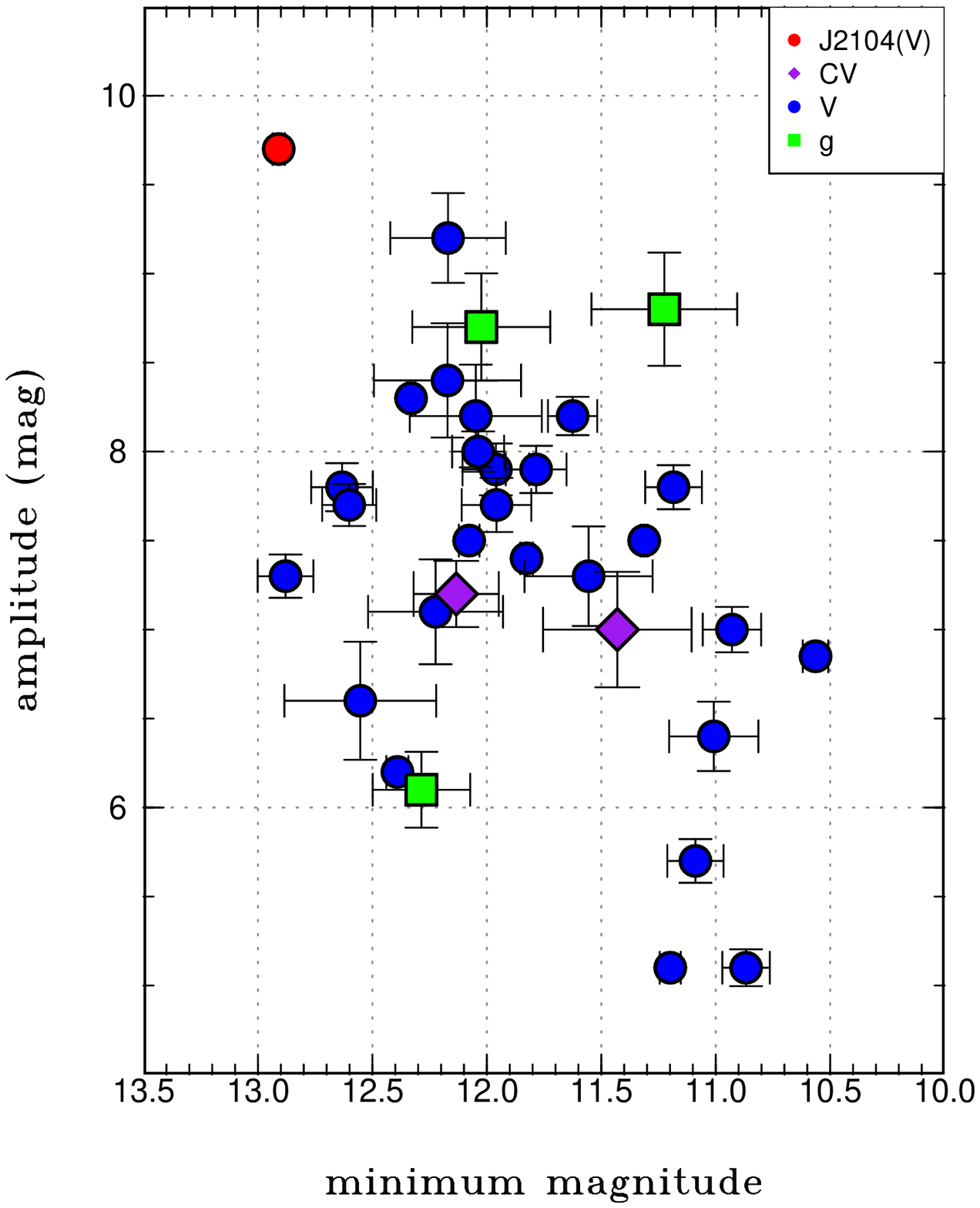}
 \end{center}
 \caption{Quiescence absolute magnitudes vs. peak absolute magnitudes of superoutbursts (left figure), and vs. superoutburst amplitudes (right figure)  of WZ Sge-type DNe from the VSX \citep{VSX} and {\it Gaia} DR 2 \citep{GaiaDR2}. Red circle represents those of TCP J2104 in the $V$ band. Other markers represent those of other WZ Sge-type DNe, and the colors and shapes of markers correspond to the filters with which the superoutburst magnitudes observed (Purple diamonds : without filter, blue circles : the $V$ band, green squares : the $g$ band). Note that if $|g - r| \ll 1$, the difference of filters can be ignored.}
 \label{fig:13}
\end{figure}

Figure \ref{fig:13} represents the relationship between absolute magnitudes in quiescence vs. those at the peaks of superoutbursts (left), and vs. the amplitudes of superoutbursts (right) of WZ Sge-type DNe. The magnitude data was taken from The International Variable Star Index (VSX; \cite{VSX}), and the distance data were taken from {\it Gaia} DR2 \citep{GaiaDR2, bai18gaia_dist}. 
Note that in Figure \ref{fig:13} only WZ Sge-type DNe with distance errors less that 20$\%$ are plotted. 
As TCP J2104 clearly showed the early superhumps and possible orbital period variations, this object is not a low-inclination system. Even through that, TCP J2104 showed one of the brightest superoutbursts and largest amplitudes among WZ Sge-type DNe.
Since the luminosity source of outbursts of DNe is the release of the gravitational energy from the accreting mass on the primary WD, this large amplitude can be accomplished by the highly accumulated mass in the accretion disk of TCP J2104 before the main outburst.

\subsection{two superoutbursts during the rebrightening phase}
\label{sec:4.3}

In this subsection, we discuss how the superoutbursts and growths of superhumps during rebrightening phases were triggered.
Essentially, in order to let the superhumps grow during the rebrightening phase, it is required for the accretion disk to expand beyond the 3:1 resonance radius again.

One possible mass supplier during the rebrightening phase is the enhanced mass transfer from the illuminated secondary \citep{ham00DNirradiation}. 
Because of the very bright superoutburst of TCP J2104, the secondary of TCP J2104 might be irradiated and heated more than other DNe, and thus, the higher enhanced mass transfer rate of TCP J2104 during the rebrightening phase might be expected.
Note that since the mass transfer rates from the secondary of SU UMa-type DNe are roughly five times higher than those of WZ Sge-type DNe \citep{hel01book}, in order to trigger two superoutbursts in half a year, the enhancement of the mass transfer rate during the rebrightening phase of TCP J2104 would be at least five times than in quiescence.
The other is the mass reservoir which is the gas left over around the outer disk.
The spectra TCP J2104 during the post-superoutburst stage showed the possible Na D feature, which was proposed to arise from a cool mass reservoir in the outer disk \citep{neu17j1222}. This mass reservoir may enable the accretion disk to grow again rapidly during the post-superoutburst stage. 
In the case of TCP J2104, the accretion disk of TCP J2104 possibly had contained more mass before the main superoutburst, and the tidal force was as weak as other WZ Sge-type DNe around the period minimum. 
These points supports more mass reservoir of TCP J2104 than other WZ Sge-type DNe. 
Note that as the amount of the mass reservoir does not increase during the rebrightening, there should be enough mass in the mass reservoir to trigger two superoutbursts right after the main superoutburst.
Compared to that, the enhanced mass transfer can sustain a mass supply from the secondary during the rebrightening phase, though the expected enhancement of the mass-transfer rate is relatively large and some numerical simulations disfavor this model as the scenario of rebrightenings \citep{osa03DNoutburst, osa04EMT}. 
Also from the point of the angular momentum, the more angular momentum the mass supplier has, the more easily the disk can expand and rebrightenings can be triggered.
The mass reservoir has that orbiting around the outer disk. On the other hand, the enhanced mass transfer has that around the circularisation radius of the system, which is only $\sim0.24A$ ($A$: binary separation; \cite{hel01book}) and this would shrink the disk rather than expand.
In fact, there is no numerical simulation code that can reproduce the rebrightening phenomena reasonably and we cannot exclude the either possible mass supplier.

From Equation \ref{eq:1} using the period of the late-stage superhumps ($P_{\rm late~stage~sh}$), the orbital period ($P_{\rm orb}$) and the mass ratio of the system ($q$), a size of the precessing disk ($r_{\rm post}$) can be calculated in Equation \ref{eq:7} \citep{kat13j1222}.

\begin{equation}
    \label{eq:7}
    1 - \frac{P_{\rm orb}}{P_{\rm late~stage~sh}} = Q(q) \times R(r_{\rm post})
\end{equation}

The estimated disk radius of TCP J2104 before the third rebrightening was 0.38(2)$A$ ($A$: binary separation; \cite{kat13qfromstageA}). 
This value is much larger than the circularisation radius of the system.
More over, after the main superoutburst, TCP J2104 dropped to $\sim$ 14 mag, however this is $\sim$ 4 mag brighter than in quiescence, and is also brighter than other WZ Sge-type DNe during their post-superoutburst stage \citep{mey15suumareb}.
As pointed by \citet{osa97egcnc, osa01egcnc, mey15suumareb}, even after the cooling wave propagates through the disk, the viscosity of the disk can be enhanced and the disk remains hotter than in quiescence. 
Note that this brighter state than in quiescence during the rebrightening phase may be attributed to the cooling of the heated WD due to the enhanced mass accretion during the main superoutburst. 
A blackbody fit to the multi-band observations by MuSCAT and MuSCAT2 during the rebrightening phase, however, yielded a size ($>10^{10}$ cm) much larger than the WD ($\sim10^8$ cm) and the disk component should have been contributed significantly.
This hotter state and large radius of the accretion disk kept TCP J2104 brighter than in quiescence and made rebrightenings easier to be triggered. 
During this phase, the existence of the late-stage superhump indicates that the disk still kept the non-axisymmetric shape after the main superoutburst, which would also make rebrightenings to be induced easier.
In additional to that, the interval between the end of main superoutburst and the first rebrightening, and between the first and second rebrightening was both $\sim$5 d, however, the interval between the second rebrightening and the start of the third rebrightening was $\sim$11 d. Around BJD 2458715, a slight brightening was observed, though this brightening did not developed to be an outburst. The rising rate of this brightening ($\sim1$ mag/d) is slower than the first, second and fourth rebrightening ($>2.5$ mag/d). 
This result suggests the brightening on BJD 2458715 was a inside-out rebrightening \citep{hel01book}.
During this slight brightening, the heating wave did not propagate through the entire disk so that the mass and angular momentum were not dissipated enough to cause an outburst but kept in the accretion disk.
This phenomenon and longer interval may enable more mass accretion on the accretion disk before the third rebrightening.

As we discussed above, the rebrightenings are expected to be more easily triggered in TCP J2104 because of the large accretion disk, presence of late-stage superhumps and more enhanced viscosity of the accretion disk compared to other WZ Sge-type DNe. 
Along with that, the more mass supply let the accretion disk be massive enough to reach the 3:1 resonance radius and initiate a superoutburst during the third rebrightening. The possible supplier are the mass reservoir and/or the transferred mass from the illuminated secondary.
The same scenario can be applied for the fifth rebrightening. After the fourth rebrightening, TCP J2104 remains $\sim$15 mag, which indicates that the accretion disk was still remained enhanced-viscosity and hotter state than in quiescence. During the long interval of $\sim$85 d, the disk became massive and finally the fifth rebrightening and the superoutburst were induced again.

\section{Summary}
\label{sec:5}
We report photometric and spectroscopic observations and analysis of the main superoutburst and five rebrightenings of TCP J21040470+4631129, a WZ Sge-type dwarf nova with multiple rebrightenings. 
The early superhump and Stage A superhump period were detected as 0.053472(5) and 0.0551(2) d, respectably.
The mass ratio of this system was estimated as 0.0880(9) using the superhump period, which is within the normal range of WZ Sge-type DNe. 
The spectra of TCP J2104 during the post-superoutburst stage showed emission and absorption lines of Balmer, He I and possible Na D. Na D feature is uncommon among dwarf novae, and suggested to imply a cool mass reservoir in the outer disk.
The slow brightening before the superoutburst and a large superoutburst amplitude suggest more accumulated mass in the accretion disk than normal WZ Sge-type DNe before the superoutburst.
In addition, we found a unique series of rebrightenings including superoutbursts and growing superhumps during the third and fifth rebrightenings, which was first ever detected during the rebrightening phase among WZ Sge-type DNe with multiple rebrightenings. 
These phenomena require the accretion disk to expand beyond the 3:1 resonance radius again during the rebrightening phase. 
The elevated brightness of the system after the main superoutburst and the presence of the late-stage superhumps suggest the enhanced viscosity and large radius of the accretion disk, which enable the rebrightenings to be triggered easier. 
Also, the massive mass reservoir because of the small mass ratio of this object and/or the transferred mass from the illuminated secondary let the accretion disk to be massive enough to initiate the superoutbursts during the rebrightening phase.

\begin{ack}

We are grateful to AAVSO, VSOLJ (especially, H. Itoh, K. Hirosawa, K. Kanatsu, S. Kiyota, K. Yoshimoto, M. Hiraga, M. Mizutani, M. Moriyama, M. Sato, M. Yamamoto) and world wide observers for providing photometric data of TCP J2104 and cataclysmic variables.
This article is partly based on observations made with the MuSCAT2 instrument, developed by ABC, at Telescopio Carlos Sánchez operated on the island of Tenerife by the IAC in the Spanish Observatorio del Teide.
The author (N. Narita) is supported by JSPS KAKENHI Grant Numbers JP18H01265 and JP18H05439, and JST PRESTO Grant Number JPMJPR1775.
The author (M. Tamura) is supported by MEXT/JSPS KAKENHI grant Nos. 18H05442, 15H02063, and 22000005.
The authors (P. A. Dubovsky, T. Medulka and I. Kudzej) acknowledge support by NSF AST-1751874 and Cottrell scholarship from the Research Corporation for Science Advancement and by the Slovak Research and Development Agency under the contract No. APVV-15-0458.
The authors (A. Zubareva, A. Belinski, A. Dodin, M. Burlak, N. Ikonnikova, E. Mishin and S. Potanin) acknowledge the support from the Program of development of M.V. Lomonosov Moscow State University (Leading Scientific School 'Physics of stars, relativistic objects and galaxies').
The authors (E. P. Pavlenko,  O. I. Antonyuk and Ju. V. Bbabina)
acknowledge support by  the RSF grant 19-72-10063.
We thank the MSU Observatory Research Program participants and especially Dr. Elias Aydi, Nathaniel Berry, Matt Bundas, Christina Conner, Alessandro Dellarovere, Hannah Gallamore, Mira Ghazali, Ben Holstad, Jessie Miller, Omid Noroozi, Shivang Patel, Barrett Ross, Courtney Wicklund, Yuzhou Wu, Yihao Zhou, Evan Zobel for their valuable assistance at the MSU Campus Observatory.
We are also thankful to the survey project ASAS-SN, {\it Gaia} and Zwicky Transient Facility for their public data sets.
We acknowledge ESA Gaia, DPAC and the Photometric Science Alerts Team (http://gsaweb.ast.cam.ac.uk/alerts)
This work has made use of data from the European Space Agency (ESA) mission {\it Gaia} (https://www.cosmos.esa.int/gaia), processed by the {\it Gaia} Data Processing and Analysis Consortium (DPAC, https://www.cosmos.esa.int/web/gaia/dpac/consortium). Funding for the DPAC has been provided by national institutions, in particular the institutions participating in the {\it Gaia} Multilateral Agreement.
Based on observations obtained with the Samuel Oschin 48-inch Telescope at the Palomar Observatory as part of the Zwicky Transient Facility project. ZTF is supported by the National Science Foundation under Grant No. AST-1440341 and a collaboration including Caltech, IPAC, the Weizmann Institute for Science, the Oskar Klein Center at Stockholm University, the University of Maryland, the University of Washington, Deutsches Elektronen-Synchrotron and Humboldt University, Los Alamos National Laboratories, the TANGO Consortium of Taiwan, the University of Wisconsin at Milwaukee, and Lawrence Berkeley National Laboratories. Operations are conducted by COO, IPAC, and UW.

\end{ack}

\section*{Supporting Information}
The following Supporting Information is available on the online version of this article: Table E1, Table E2, Table E3, Figure E1, Figure E2, and Figure E3.


\bibliographystyle{pasjtest1}
\bibliography{cvs}

\newcommand{\noop}[1]{}
\begin{thebibliography}{}

\bibitem[{Abril} et~al.(2020)]{abr20CVinGaiaHRD}
  {Abril}, Javier, {Schmidtobreick}, Linda, {Ederoclite}, Alessand~ro, \&
  {L{\'o}pez-Sanjuan}, Carlos\ 2020, \mnras, 492, L40

\bibitem[{Bailer-Jones} et~al.(2018)]{bai18gaia_dist}
  {Bailer-Jones}, C.~A.~L., {Rybizki}, J., {Fouesneau}, M., {Mantelet}, G., \&
  {Andrae}, R.\ 2018, \aj, 156, 58

\bibitem[{Bellm} et~al.(2019)]{ZTF}
  {Bellm}, Eric~C., {et~al.}\ 2019, PASP, 131, 018002

\bibitem[Clarke and Bowyer(1984)]{cla84rxandktper}
  Clarke, J.~T., \& Bowyer, S.\ 1984, A\&A, 140, 345

\bibitem[{Cleveland}(1979)]{LOWESS}
  {Cleveland}, W.~S.\ 1979, J. Amer. Statist. Assoc., 74, 829

\bibitem[Fernie(1989)]{fer89error}
  Fernie, J.~D.\ 1989, PASP, 101, 225

\bibitem[{Gaia Collaboration} et~al.(2018)]{GaiaDR2}
  {Gaia Collaboration}, {et~al.}\ 2018, \aap, 616, A1

\bibitem[{Gaia Collaboration} et~al.(2016)]{gaia}
  {Gaia Collaboration}, {et~al.}\ 2016, A\&A, 595, A1

\bibitem[{Green} et~al.(2019)]{gre19dustextinction}
  {Green}, Gregory~M., {Schlafly}, Edward~F., {Zucker}, Catherine, {Speagle},
  Joshua~S., \& {Finkbeiner}, Douglas~P.\ 2019, arXiv e-prints,
  arXiv:1905.02734

\bibitem[Hameury et~al.(2000)]{ham00DNirradiation}
  Hameury, J.-M., Lasota, J.-P., \& Warner, B.\ 2000, A\&A, 353, 244

\bibitem[Hellier(2001)]{hel01book}
  Hellier, C.\ 2001, Cataclysmic Variable Stars: How and why they vary (Berlin:
  Springer)

\bibitem[{Hirose} and {Osaki}(1990)]{hir90SHexcess}
  {Hirose}, M., \& {Osaki}, Y.\ 1990, PASJ, 42, 135

\bibitem[{Imada} et~al.(2006)]{ima06tss0222}
  {Imada}, A., {Kubota}, K., {Kato}, T., {Nogami}, D., {Maehara}, H.,
  {Nakajima}, K., {Uemura}, M., \& {Ishioka}, R.\ 2006, PASJ, 58, L23

\bibitem[Ishioka et~al.(2002)]{ish02wzsgeproc}
  Ishioka, R., Uemura, M., Kato, T., \& {The VSNET Collaboration Team}\ 2002,
  in ASP\ Conf.\ Ser.\ 261, The Physics of Cataclysmic Variables and Related
  Objects, ed. B.~T. {G\"ansicke}, K. Beuermann, \& K. Reinsch (San Francisco:
  ASP), p.~491

\bibitem[{Isogai} et~al.(2019)]{iso19asassn14dx}
  {Isogai}, K., {et~al.}\ 2019, PASJ, 71, 22

\bibitem[{Isogai} et~al.(2015)]{iso15ezlyn}
  {Isogai}, M., {Arai}, A., {Yonehara}, A., {Kawakita}, H., {Uemura}, M., \&
  {Nogami}, D.\ 2015, PASJ, 67, 7

\bibitem[Kato(2002)]{kat02wzsgeESH}
  Kato, T.\ 2002, PASJ, 54, L11

\bibitem[{Kato}(2015)]{kat15wzsge}
  {Kato}, T.\ 2015, PASJ, 67, 108

\bibitem[{Kato} et~al.(2009)]{Pdot}
  {Kato}, T., {et~al.}\ 2009, PASJ, 61, S395

\bibitem[{Kato} et~al.(2016)]{kat16rzlmi}
  {Kato}, T., {et~al.}\ 2016, PASJ, 68, 107

\bibitem[{Kato} et~al.(2017)]{Pdot9}
  {Kato}, T., {et~al.}\ 2017, PASJ, 69, 75

\bibitem[{Kato} et~al.(2010)]{Pdot2}
  {Kato}, T., {et~al.}\ 2010, PASJ, 62, 1525

\bibitem[{Kato} et~al.(2013)]{kat13j1222}
  {Kato}, T., {Monard}, B., {Hambsch}, F.-J., {Kiyota}, S., \& {Maehara}, H.\
  2013, PASJ, 65, L11

\bibitem[Kato et~al.(1997)]{kat97egcnc}
  Kato, T., Nogami, D., Matsumoto, K., \& Baba, H.\ 1997,
  ftp://vsnet.kusastro.kyoto-u.ac.jp/pub/vsnet/preprints/EG\_Cnc/

\bibitem[Kato et~al.(2004a)]{kat04egcnc}
  Kato, T., Nogami, D., Matsumoto, K., \& Baba, H.\ 2004a, PASJ, 56, S109

\bibitem[{Kato} and {Osaki}(2013)]{kat13qfromstageA}
  {Kato}, T., \& {Osaki}, Y.\ 2013, PASJ, 65, 115

\bibitem[Kato et~al.(2004b)]{VSNET}
  Kato, T., Uemura, M., Ishioka, R., Nogami, D., Kunjaya, C., Baba, H., \&
  Yamaoka, H.\ 2004b, PASJ, 56, S1

\bibitem[{Knigge} et~al.(2011)]{kni11CVdonor}
  {Knigge}, C., {Baraffe}, I., \& {Patterson}, J.\ 2011, ApJS, 194, 28

\bibitem[{Kuulkers} et~al.(2011)]{kuu11wzsge}
  {Kuulkers}, E., {Henden}, A.~A., {Honeycutt}, R.~K., {Skidmore}, W.,
  {Waagen}, E.~O., \& {Wynn}, G.~A.\ 2011, A\&A, 528, A152

\bibitem[Lin and Papaloizou(1979)]{lin79lowqdisk}
  Lin, D. N.~C., \& Papaloizou, J.\ 1979, MNRAS, 186, 799

\bibitem[{Lubow}(1991)]{lub91SHa}
  {Lubow}, S.~H.\ 1991, ApJ, 381, 259

\bibitem[Masci et~al.(2018)]{ztf_cite}
  Masci, Frank~J., {et~al.}\ 2018, Publications of the Astronomical Society of
  the Pacific, 131, 018003

\bibitem[{Matsubayashi} et~al.(2019)]{mat19koolsifu}
  {Matsubayashi}, Kazuya, {et~al.}\ 2019, PASJ, 71, 102

\bibitem[Meyer and Meyer-Hofmeister(1981)]{mey81DNoutburst}
  Meyer, F., \& Meyer-Hofmeister, E.\ 1981, A\&A, 104, L10

\bibitem[{Meyer} and {Meyer-Hofmeister}(2015)]{mey15suumareb}
  {Meyer}, F., \& {Meyer-Hofmeister}, E.\ 2015, PASJ, 67, 52

\bibitem[{Narita} et~al.(2015)]{MuSCAT}
  {Narita}, Norio, {et~al.}\ 2015, Journal of Astronomical Telescopes,
  Instruments, and Systems, 1, 045001

\bibitem[{Narita} et~al.(2019)]{MuSCAT2}
  {Narita}, Norio, {et~al.}\ 2019, Journal of Astronomical Telescopes,
  Instruments, and Systems, 5, 015001

\bibitem[{Neustroev} et~al.(2019a)]{neu19j2104orb}
  {Neustroev}, V., {et~al.}\ 2019a, The Astronomer's Telegram, 13009, 1

\bibitem[{Neustroev} et~al.(2019b)]{19neuatelj2104}
  {Neustroev}, V., {et~al.}\ 2019b, The Astronomer's Telegram, 13297, 1

\bibitem[{Neustroev} et~al.(2017)]{neu17j1222}
  {Neustroev}, V.~V., {et~al.}\ 2017, MNRAS, 467, 597

\bibitem[{Osaki}(1974)]{osa74DNmodel}
  {Osaki}, Y.\ 1974, PASJ, 26, 429

\bibitem[{Osaki}(1989)]{osa89suuma}
  {Osaki}, Y.\ 1989, PASJ, 41, 1005

\bibitem[{Osaki} and {Meyer}(2002)]{osa02wzsgehump}
  {Osaki}, Y., \& {Meyer}, F.\ 2002, A\&A, 383, 574

\bibitem[{Osaki} and {Meyer}(2003)]{osa03DNoutburst}
  {Osaki}, Y., \& {Meyer}, F.\ 2003, A\&A, 401, 325

\bibitem[{Osaki} and {Meyer}(2004)]{osa04EMT}
  {Osaki}, Y., \& {Meyer}, F.\ 2004, A\&A, 428, L17

\bibitem[{Osaki} et~al.(2001)]{osa01egcnc}
  {Osaki}, Y., {Meyer}, F., \& {Meyer-Hofmeister}, E.\ 2001, A\&A, 370, 488

\bibitem[{Osaki} et~al.(1997)]{osa97egcnc}
  {Osaki}, Y., {Shimizu}, S., \& {Tsugawa}, M.\ 1997, PASJ, 49, L19

\bibitem[{Patterson} et~al.(2005)]{pat05SH}
  {Patterson}, J., {et~al.}\ 2005, PASP, 117, 1204

\bibitem[{Pavlenko} et~al.(2019)]{pav19a19fk}
  {Pavlenko}, E., {et~al.}\ 2019, Contributions of the Astronomical Observatory
  Skalnate Pleso, 49, 204

\bibitem[Richter(1992)]{ric92wzsgedip}
  Richter, G.~A.\ 1992, in ASP\ Conf.\ Ser.\ 29, Vi{\~n}a del Mar Workshop on
  Cataclysmic Variable Stars, ed. N. Vogt (San Francisco: ASP), p.~12

\bibitem[{Scaringi}(2014)]{sca14flickering}
  {Scaringi}, S.\ 2014, MNRAS, 438, 1233

\bibitem[{Shappee} et~al.(2014)]{ASASSN}
  {Shappee}, B.~J., {et~al.}\ 2014, ApJ, 788, 48

\bibitem[{Sokolovsky} et~al.(2019)]{neu19j2104-2}
  {Sokolovsky}, K., {et~al.}\ 2019, The Astronomer's Telegram, 12947, 1

\bibitem[Stellingwerf(1978)]{PDM}
  Stellingwerf, R.~F.\ 1978, ApJ, 224, 953

\bibitem[{Uemura} et~al.(2008)]{uem08j1021}
  {Uemura}, M., {et~al.}\ 2008, PASJ, 60, 227

\bibitem[{Uemura} et~al.(2012)]{uem12ESHrecon}
  {Uemura}, M., {Kato}, T., {Ohshima}, T., \& {Maehara}, H.\ 2012, PASJ, 64, 92

\bibitem[{van Spaandonk} et~al.(2010)]{vanspa10gwlib}
  {van Spaandonk}, L., {Steeghs}, D., {Marsh}, T.~R., \& {Torres}, M.~A.~P.\
  2010, MNRAS, 401, 1857

\bibitem[Warner and Woudt(2002)]{war02DNO}
  Warner, B., \& Woudt, P.~A.\ 2002, MNRAS, 335, 84

\bibitem[{Watson} et~al.(2006)]{VSX}
  {Watson}, C.~L., {Henden}, A.~A., \& {Price}, A.\ 2006, Society for
  Astronomical Sciences Annual Symposium, 25, 47

\bibitem[Whitehurst(1988)]{whi88tidal}
  Whitehurst, R.\ 1988, MNRAS, 232, 35

\end{thebibliography}


\appendix

\end{document}